# Co-evolution is Incompatible with the Markov Assumption in Phylogenetics

Tamir Tuller[1,2] and Elchanan Mossel[1,3]

[1]Faculty of Mathematics and Computer Science, [2] Department of Molecular Genetics, Weizmann Institute of Science. Rehovot, Israel; [3] Departments of Statistics and Computer Science, U.C. Berkeley.

**Markov models are extensively used in the analysis of molecular evolution. A recent line of research suggests that pairs of proteins with functional and physical interactions co-evolve with each other. Here, by analyzing hundreds of orthologous sets of three fungi and their co-evolutionary relations, we demonstrate that co-evolutionary assumption may violate the Markov assumption. Our results encourage developing alternative probabilistic models for the cases of extreme co-evolution.**

Markov models have been extensively used in studies and modeling of molecular evolution (see, for example, [1-11]). The Markov assumption is very natural: stating that the statistical distribution of nucleotides in different positions of a gene is determined by their distribution in the corresponding gene of its direct ancestor with no effect of older ancestors (**Figure 1A**).

Let $X_i^k$ denote a random variables corresponding to the value of the *k*-th nucleotide at a given site at node (taxa) *i*. Let $x_i^k$ denote a particular state (nucleotide) of the random variable $X_i^k$.

By the Markov assumption, if $X_1^k$ is the direct ancestor of $X_2^k$, and $X_2^k$ is the direct ancestor of $X_3^k$, the following is true:

$$p(X_3^k \mid X_2^k, X_1^k) = p(X_3^k \mid X_2^k)$$

We say that two sites, $X_i^{k1}$ *and* $Y_i^{k2}$, co-evolve if evolutionary change at one site is influenced by the evolution of the second site. Mathematically it means that there is information flow between $X_i^{k1}$ and $Y_i^{k2}$ (*i.e.* the joint distribution of $X_i^{k1}$ [or $Y_i^{k2}$] depends on both $X_{i-1}^{k1}$ and $Y_{i-1}^{k2}$; see **Figure 1B**).
A recent line of research suggests that different sites within or between proteins functionally and physically interact and thus, co-evolve (see, for example, [12-17]).

As can be seen in **Figure 1B**, if site $X_i^{k1}$ *co-evolves with site* $Y_i^{k2}$ there may be an information flow between $X_3^{k1}$ and its indirect ancestor $X_1^{k1}$, **not via** $X_2^{k1}$. Thus, mathematically, under such a realistic assumption, a site in a certain node (taxon) in the evolutionary tree may depend on the value of its corresponding indirect ancestors even when conditioning on its direct ancestor, contradicting the Markov assumption (**Figure 1B**).

To demonstrate this point, we analyzed the conserved coding sequences of three close fungi (Figure 1C; Methods); we aimed at performing a statistical test that checks the Markov assumption without any additional assumptions on the nature or parameters of the process. Specifically, we computed a measurement that is related to $[p(X_3^k | X_2^k, X_1^k) - p(X_3^k | X_2^k)] / p(X_3^k | X_2^k)$ (*i.e.* it measures the relative skew from the Markov assumption considering three nodes in the evolutionary tree; see more details in the Methods).

We found a significant positive relation between the skew from the Markov assumption and the density of co-evolutionary relations (the number of co-evolutionary relations of a gene normalized by its length; see technical details in the Methods) - more co-evolutionary interactions per nucleotide implies larger skew from the Markov assumption. Specifically, when we compared the 15% of the genes with top co-evolutionary density to the 15% of the genes with the bottom co-evolutionary density we found that the first group has significantly higher mean skew from the Markov assumption (T-test – p value = $6.5*10^{-5}$, KS test p value = $3*10^{-6}$). In addition, we found significant Spearman correlation between co-evolutionary densities and skew from the Markov assumption across all genes (r= 0.15, p = $2.8*10^{-4}$; Spearman correlation, 10 bins with equal size, each with 10% of the genes: r= 0.84, p = 0.002; **Figure 1C**). The correlation remained significant even when we controlled for the conservation of the genes (r = 0.134; p=0.001; Methods) demonstrating that different mutation rate between genes can not explain the correlation.

Our results suggest that co-evolution introduces memory to the process of molecular evolution. Moreover, the density of co-evolutionary relations of a gene is inversely related to how well a Markov model approximates its evolution.

Previous studies have shown (based on simulation and analytical analysis) that skew from Markovity can cause erroneous phylogenetic reconstruction [18, 19] and increase the error rate in ancestral reconstruction [12]. Thus, we should expect higher error rate when we use Markovian models to analyzed genomic sequences that have many co-evolutionary relations. In addition, our results encourage developing/using alternative probabilistic models for the cases of extreme co-evolution; one possible alternative probabilistic model might be a hidden Markov model (**Figure 1D**) where the hidden variables represent the interaction between the protein/site and other proteins/sites.

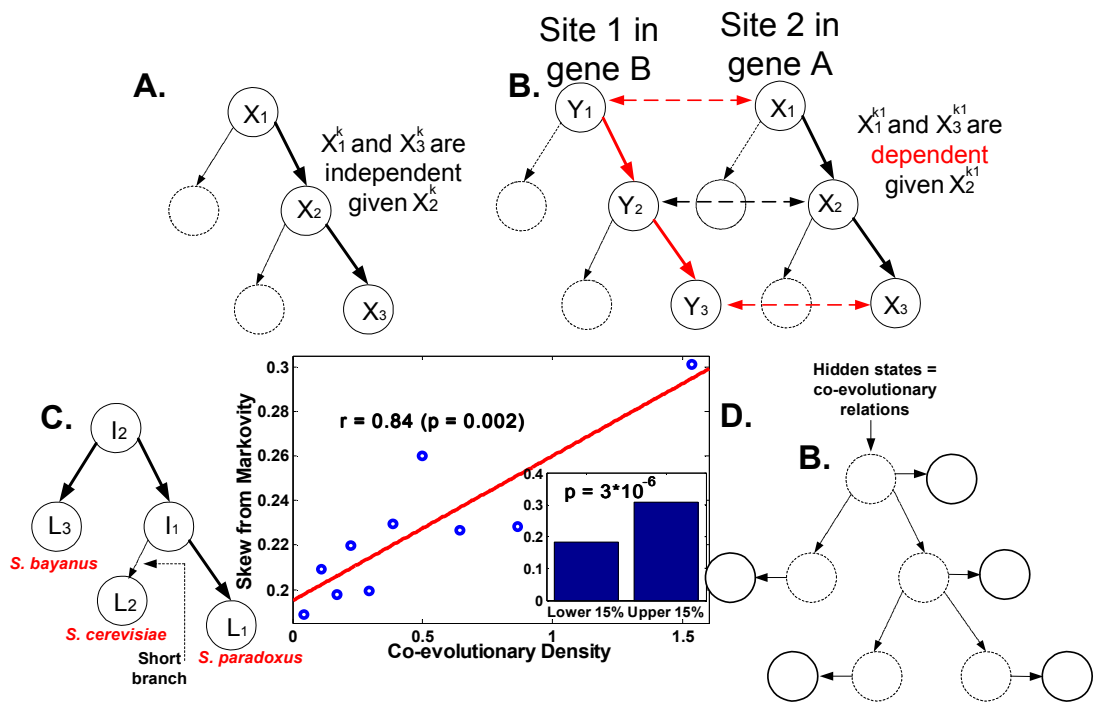

**Figure 1. A.** The traditional model of molecular evolution is Markovian; each arrow represents the information flow from an ancestor to its descendant. The value of a node does not depend on its 'grandparent' given its direct parent. **B.** A model of molecular evolution under co-evolution; the dashed arrows represent co-evolutionary relations. Red arrows are used to show the route by which information may 'flow around' an immediate ancestor. The fact that two proteins/sites co-evolve and thus they are dependent implies that the value of a node may depend on its grandparent given its direct parent. **C.** The skew from Markovity, measured by analyzing the coding sequences of three fungi, increases with the density of co-evolutionary relations (10 bins of equal size, 10% of the genes, of co-evolutionary density vs. the skew from Markovity); the correlation between the mean co-evolutionary density and Markovity is significantly higher for the 15% of the genes with the highest co-evolutionary density compared to the 15% of the genes with the lowest co-evolutionary density (KS test p value = $3*10^{-6}$) . **D.** An illustration of probabilistic model that may better describe the evolution of a single site or a protein under extensive co-evolution.

Finally, it is important to mention that co-evolution is not the only possible cause of non-Markov behavior. For example, it was suggested before that when the substitution rates vary across sites the entire probabilistic process becomes non-Markovian [18]. It is easy to see that co-evolution and varying substitution rates are not independent phenomena (Figure 2): proteins that physically interact with each other tend to co-evolve ( [12, 13, 20] ; Figure 2A); in these proteins, the sites that are involved in the interactions are expected to have less substitutions as they are under more constraints (Figure 2B). Thus, co-evolution can induce varying substitution rates. There are many additional possible reasons that may cause a skew from

Markovity; some of them are the functionalities of different parts of the proteins (that may have different substitution rates), the position within a codon (it is known that the third positions are less conserved [18]), the fact that different regions (*e.g.* the beginning of the coding region [21, 22]) correspond to the regulation of its translation [21, 22] and thus may have different substitution rate.

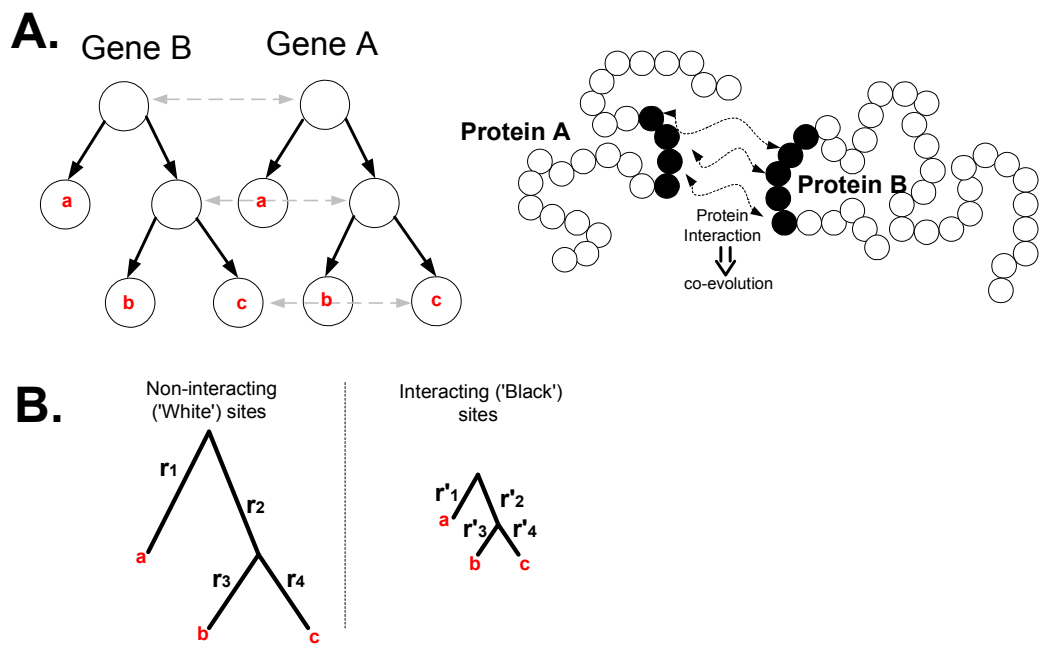

**Figure 2. An illustration describing how co-evolution can imply varying substitution rate across characters. A. The proteins A and B physically interact with each other and thus they co-evolve. The interacting sites within each protein are marked in black. B. The substitution rates in the non interacting sites ($r_1$,..,$r_4$) and in the interacting sites ($r'_1$,..,$r'_4$) of protein A – the non interacting sites are under less evolutionary constraints and thus have higher substitution rate.**

**Methods**:

**Sets of orthologs:** The sequences of the three Fungi (*S. paradoxus*, *S. cerevisiae*, and *S. bayanus*) and mapping of genes to groups of orthologs were downloaded from [23]. We considered sets of orthologs that do not include duplicates (according to COG [24]) and whose level of conservation (percentage of the sites that are identical in the three Fungi) is above 30%. The final dataset included 598 sets of orthologs

**Alignment:** We aligned each set (of three coding sequences) using CLUSTALW [25]. Specifically, for each set, we translated the three sequences to Amino Acids; align them, removed gaps, and converted the result amino acid sequences to an aligned set of nucleotide sequences.

**Co-evolution relations:** The co-evolutionary information (based on a composite score that is based on co-expression, co-occurrences in the same genome, genomic proximity, protein-protein interaction, and more) was downloaded from STRING [26]. We mapped the *S. cerevisiae* gene in each set to a corresponding COG; the number of relations of the COG in STRING was used as an estimator of the level of co-evolution for the set of orthologs. To compute the *density* of the co-evolutionary relations we divided this number by the length of the alignment.

**The phylogenetic tree:** We used the phylogentic tree of [27].

**Estimating the values at the ancestors of *S. paradoxus*:** To study how the Markov property relates to co-evolution we are interested in Markov chains of length three (*e.g. S. paradoxus*, the direct ancestor of *S. paradoxus*, and *S. bayanus* – that was used as an indirect ancestor of *S. paradoxus* under Bayes rule; **Figure 1C**). In our analysis, we do not know the actual values at the direct ancestor of *S. paradoxus*. However, the branch connecting *S. cerevisiae* to the ancestor of *S. paradoxus* is relatively short ( the edge length is 0.015; see [27]): it is more than 3 times shorter than the other branches to the leaves in the analyzed tree; it is also much shorter (at least 3 times shorter, many times more than 7 times shorter) than all the branches to the 42 Fungi that appear in the original tree (see Figure 2 in [27]).

Thus, we used the value at the genome of *S. cerevisiae* as an estimator for the values at the direct ancestors of *S. paradoxus*. It is important to emphasize that in this paper we show that our measure of skew from the Markov property (that is based on the assumption that the edge length above is very short and thus may be noisy) is correlated with the density of co-evolutionary relations. This correlation can not be explained by the by noise in our measure (if the noise is not related to co-evolution). In addition, based on Bayes' law (or assume a reversible stochastic model, see, for example, [28]) we use *S. bayanus* as indirect ancestor of *S. paradoxus*. Note that the same assumptions and approximations were made for all the genes (*i.e.* both for genes with high density of co-evolutionary relations and the genes with low density of co-evolutionary relations).

**Checking for the Markov property:** We design a statistical test to estimate the Markov property; the test does not require any additional assumptions on the nature or parameters of the process.

For a certain set of orthologs, let $x_3^{k1}, x_3^{k2}$ denote the values at two sites in a gene of *S. paradoxus*; let $x_2^{k1}, x_2^{k2}$ denote the values at the corresponding sites at the direct ancestor of *S. paradoxus*, let $x_1^{k1}, x_1^{k2}$ denote the value at the corresponding sites of the indirect ancestor of *S. paradoxus*.

Under the Markov property, the statistical distribution of nucleotides in different positions of a gene is determined by their distribution in the corresponding gene of its direct ancestor with no effect of older ancestors. In our case, we assume only the three organism mentioned above [the organisms for which the required data were available and which satisfied our assumption about the edge lengths].

We aimed at comparing

(1) $\Pr[x_3^{k1} = x_3^{k2} \mid x_2^{k1} = x_2^{k2}]$ and (2) $\Pr[x_3^{k1} = x_3^{k2} \mid x_2^{k1} = x_2^{k2}, x_1^{k1} = x_1^{k2}]$.

For a Markov model we expect that (1) will be equal to (2) but for a non-Markovian case we expect that (2) will be larger than (1).

Let $\delta(\cdot)$ denote the indicator function; for a certain gene, these values were estimated by the following equations:

First, we considered all the pairs of sites that are identical at the direct ancestral gene and computed the fraction of times that the corresponding pair of sites at the gene is also identical (*i.e.* this is the empirical probability that a pair of sites are identical given that they are identical at the direct ancestral gene).

(3) $I_{3,2} \equiv \dfrac{\sum_{k1,k2} \delta(x_3^{k1} = x_3^{k2}, x_2^{k1} = x_2^{k2})}{\sum_{k1,k2} \delta(x_2^{k1} = x_2^{k2})}$

Second, we considered all pairs of sites that are identical both in the direct ancestral gene and in the indirect ancestral gene and computed the number of times the corresponding pair of sites at the gene is also identical (*i.e.* this is the empirical probability that a pair of sites are identical given that they are identical at the direct and indirect ancestral gene).

$$(4) \quad I_{3,2,1} \equiv \frac{\sum_{k1,k2} \delta(x_3^{k1} = x_3^{k2}, x_2^{k1} = x_2^{k2}, x_1^{k1} = x_1^{k2})}{\sum_{k1,k2} \delta(x_2^{k1} = x_2^{k2}, x_1^{k1} = x_1^{k2})}$$

By the markovian assumption (if the direct ancestor is known information about the indirect ancestor should not help determining the value at the current site) we do not expect that (4) will be larger than (3). Thus, we used (5) $(I_{3,2,1} - I_{3,2})/I_{3,2}$ to estimate the skew from the Markov property in each gene.

It is known that proteins with more co-evolutionary relations are more conserved [29]. Let, $|x|$ denote the length of the sequence (gene) $x$; let $C = \sum_k \delta(x_1^k = x_2^k = x_3^k)/|x|$ denote the conservation level of a gene $x$ (in our case, the fraction of the sites in x that are conserved in the three Fungi that we analyzed). In our dataset, the correlation between $C = \sum_k \delta(x_1^k = x_2^k = x_3^k)/|x|$ and the number of co-evolutionary relations is 0.135; p = 0.0008. Thus, we used $C$ as a covariate variable in the partial correlation between the density of co-evolutionary relations and the skew from Markovity.

In addition, to control for the fact that conservation of proteins with high co-evolutionary relations is higher [29] and as we are interested in the statistical nature of the phenomenon and not in its biological/functional nature we considered only pairs

$k_1$ and $k_2$ for which $x_3^{k1} = x_3^{k2}, x_2^{k1} = x_2^{k2}, x_3^{k1} \neq x_2^{k1}$ and applied this restriction to the numerator and denominators of (3) and (4).

## Acknowledgments

We thank Prof. Martin Kupiec and Prof. Eytan Ruppin for helpful discussions. We also want to thank the two anonymous reviewers and the editor, Prof. Mike Steel, for their helpful comments. T.T. is a Koshland Scholar at Weizmann Institute of Science and is supported by a travel fellowship from EU grant PIRG04-GA-2008-239317. E.M. was supported by US NSF DMS 0548249 (CAREER) award, by Israel Science Foundation grant 1300/08 and by EU grant PIRG04-GA-2008-239317.

## References

1. Felsenstein, J., *Evolutionary trees from DNA sequences: a maximum likelihood approach.* J Mol Evol, 1981. **17**(6): p. 368-76.
2. Pagel, M., *Inferring the historical patterns of biological evolution.* Nature, 1999. **401**(6756): p. 877-84.
3. Huelsenbeck, J.P. and B. Rannala, *Phylogenetic methods come of age: testing hypotheses in an evolutionary context.* Science, 1997. **276**(5310): p. 227-32.
4. Jermiin, L.S., L. Poladian, and M.A. Charleston, *Evolution. Is the "Big Bang" in animal evolution real?* Science, 2005. **310**(5756): p. 1910-1.
5. Bridgham, J.T., S.M. Carroll, and J.W. Thornton, *Evolution of hormone-receptor complexity by molecular exploitation.* Science, 2006. **312**(5770): p. 97-101.
6. Teeling, E.C., et al., *A molecular phylogeny for bats illuminates biogeography and the fossil record.* Science, 2005. **307**(5709): p. 580-4.
7. Dieterich, C., et al., *The Pristionchus pacificus genome provides a unique perspective on nematode lifestyle and parasitism.* Nat Genet, 2008. **40**(10): p. 1193-8.
8. Delsuc, F., H. Brinkmann, and H. Philippe, *Phylogenomics and the reconstruction of the tree of life.* Nat Rev Genet, 2005. **6**(5): p. 361-75.
9. Thornton, J.W., *Resurrecting ancient genes: experimental analysis of extinct molecules.* Nat Rev Genet, 2004. **5**(5): p. 366-75.
10. Chang, J.T., *Full reconstruction of Markov models on evolutionary trees: identifiability and consistency.* Math Biosci, 1996. **137**(1): p. 51-73.

11. Elias, I. and T. Tuller, *Reconstruction of ancestral genomic sequences using likelihood.* J. Comput. Biol., 2007. **14**(2): p. 216-237.
12. Tuller, T., et al., *Reconstructing ancestral gene content by coevolution.* Genome Res, 2009. **20**(1): p. 122-32.
13. Juan, D., F. Pazos, and A. Valencia, *High-confidence prediction of global interactomes based on genome-wide coevolutionary networks.* Proc Natl Acad Sci U S A, 2008. **105**(3): p. 934-9.
14. Pellegrini, M., et al., *Assigning protein functions by comparative genome analysis: protein phylogenetic profiles.* Proc Natl Acad Sci U S A, 1999. **96**(8): p. 4285-8.
15. Yeang, C.H. and D. Haussler, *Detecting coevolution in and among protein domains.* PLoS Comput Biol, 2007. **3**(11): p. e211.
16. Lockless, S.W. and R. Ranganathan, *Evolutionarily conserved pathways of energetic connectivity in protein families.* Science, 1999. **286**(5438): p. 295-9.
17. Tuller, T., Y. Felder, and M. Kupiec, *Discovering local patterns of co-evolution: computational aspects and biological examples.* BMC Bioinformatics. **11**(43): p. 43.
18. Chang, J.T., *Inconsistency of evolutionary tree topology reconstruction methods when substitution rates vary across characters.* Math Biosci, 1996. **134**(2): p. 189-215.
19. Matsen, F.A. and M. Steel, *Phylogenetic mixtures on a single tree can mimic a tree of another topology.* Syst Biol, 2007. **56**(5): p. 767-75.
20. Tuller, T., M. Kupiec, and E. Ruppin, *Co-evolutionary networks of genes and cellular processes across fungal species.* Genome Biol, 2009. **10**(5): p. R48.
21. Tuller, T., et al., *An evolutionarily conserved mechanism for controlling the efficiency of protein translation.* Cell. **141**(2): p. 344-54.
22. Tuller, T., et al., *Translation efficiency is determined by both codon bias and folding energy.* Proc Natl Acad Sci U S A. **107**(8): p. 3645-50.
23. Kellis, M., et al., *Sequencing and comparison of yeast species to identify genes and regulatory elements.* Nature, 2003. **423**(6937): p. 241-54.
24. Tatusov, R.L., et al., *The COG database: an updated version includes eukaryotes.* BMC Bioinformatics, 2003. **4**(41): p. 41.
25. Thompson, J.D., T.J. Gibson, and D.G. Higgins, *Multiple sequence alignment using ClustalW and ClustalX.* Curr Protoc Bioinformatics, 2002. **Chapter 2**(2): p. Unit 2 3.
26. Jensen, L.J., et al., *STRING 8--a global view on proteins and their functional interactions in 630 organisms.* Nucleic Acids Res, 2009. **37**(Database issue): p. D412-6.
27. Fitzpatrick, D.A., et al., *A fungal phylogeny based on 42 complete genomes derived from supertree and combined gene analysis.* BMC Evol Biol, 2006. **6**(99): p. 99.
28. Waddell, P.J. and M.A. Steel, *General time-reversible distances with unequal rates across sites: mixing gamma and inverse Gaussian distributions with invariant sites.* Mol Phylogenet Evol, 1997. **8**(3): p. 398-414.
29. Fraser, H.B., et al., *Evolutionary rate in the protein interaction network.* Science, 2002. **296**(5568): p. 750-2.